\documentclass[preprint, aps, amsmath, amssymb, nobibnotes]{revtex4-1}
\usepackage{graphicx}   
\usepackage{epstopdf}
\begin{document}

\title{High-resolution Experimental Study and Numerical Modeling of Population Dynamics in a Bacteria Culture}

\author{A. Chatterjee $^\dag$ \email[Email:]{atanu@wpi.edu}, N. Mears $^\dag$, A.J. Charest $^\ddag$, S. Algarni $^\sharp$, G.S. Iannacchione $^\dag$ \email[Email:]{gsiannac@wpi.edu} }
\affiliation{$^\sharp$ Department of Physics, King Saud University, Riyadh, Saudi Arabia}
\affiliation{$^\ddag$ Department of Civil Engineering, Wentworth Institute of Technology, Boston, MA, 02115, USA}
\affiliation{$^\dag$ Department of Physics, Worcester Polytechnic Institute, Worcester, MA, 01609, USA}

\date{\today}


\begin{abstract}
In this paper, experimental data is presented and a simple model is developed for the time evolution of a F-amp \textit{E. Coli} culture population. In general, the bacteria life cycle as revealed by monitoring a culture's population consists of the lag phase, the growth (or exponential) phase, the log (or stationary) phase, and finally the death phase. As the name suggests, in the stationary phase, the population of the bacteria ceases to grow exponentially and reaches a plateau before beginning the death phase. High temporal resolution experimental observations using a unique light-scattering technique in this work reveal all the expected phases in detail as well as an oscillatory population behavior in the stationary phase. This unambiguous oscillation behavior has been suggested previously using traditional  surveys of aliquots from a given population culture. An attempt is made to model these experimental results by developing a differential equation that accounts for the spatial distribution of the individual cells and the presence of the self-organizing forces of competition and dispersion. The main phases are well represented, and the oscillating behavior is attributed to intra-species mixing. It is also observed, that the convective motion arising out of intra-species mixing while plays a key role in limiting population growth, scales as $t^{-\alpha}$, where $\alpha$ is bounded by model parameters.
\end{abstract}

\maketitle

\section{Introduction}
The application of differential equations to model population dynamics of biological systems is not a new phenomenon. From simple, exponential population growth equations to complex systems of differential equations such as, the predator-prey model have widely been used to model the population dynamics of a wide array of biological species. The problem of finding how the population of a species or a group of different species grow and interact with each other in time, has traditionally been a dominant branch of mathematical biology. While of intense theoretical interest because of its broad implications in biology, biophysics, ecology, and even economics, monitoring population dynamics experimentally has also attracted considerable activity~\cite{Liu2017}. When focused on bacterial populations, the experimental challenges are significant, especially for high-resolution measurements over a long period of time. Typically, monitoring a bacterial population's behavior involves culture or genomic methods to monitor aliquots from a evolving bacterial culture that leads to statistical uncertainties of the representative samples as well as time-base uncertainties due to the surveys requiring an extended time to be performed. This work reports high-resolution data that is free from those mentioned uncertainties on bacterial population behavior along with developing a model to represent this behavior.

\section{Mathematical Models for Population Growth}
One of the earliest models to mathematically formulate the growth of a single species in time is the simple exponential growth model or the Malthusian model~\cite{malthus1809essay}. It is apparent that the Malthusian growth curve is not a realistic ecological growth model, as the population of a single species cannot, indefinitely, grow in time. Also, in a more realistic ecological setting there is always a competition between members of a species for space and food. Accounting for these interactions, the Malthusian growth smoothens over time and results in a logistic curve, which was proposed by Verhulst~\cite{verhulst1845recherches}. In real ecosystems there are more than one species competing against each other. Predation and prey mechanisms have been found to stabilize the population of different species into limit cycles. One of the earliest models that takes into account the predator-prey mechanism in ecosystems is that of the Lotka-Volterra competition model~\cite{bomze1983lotka}. The competition between the interacting species is denoted by a non-linear term that (\textbf{i}) limits the exponential growth and (\textbf{ii}) signifies the inter-species interaction. Although the competition model fairly well replicates real-world bio-systems, it suffers from two important drawbacks. In both the Malthusian growth and the Lotka-Volterra competition models, it is assumed that the growth and interaction occur at the instant a species encounters a food source, or a predator encounters a prey. This is however not true as there is a time delay associated with population growth and predation, usually denoted as the lag-phase. The second important drawback is the fact that the population growth, interaction and death are solely time dependent phenomena. The dispersal effects of the population and spatial dependence on interaction are completely ignored. In order to account for the above discrepancies, numerous extensions of the existing models have been proposed such as the competition model involving delay, evolutionary game theory, growth-diffusion equation signifying dispersion, stochastic dynamical equations and spatial logistic equations~\cite{mohr2013predator, van1988cannibalism, cushing2015evolutionary, law2003population, magnusson1999destabilizing}.

As discussed above, the Malthusian growth model is the simplest mathematical model describing population growth. According to this model, the population growth rate is directly proportional to the number of individuals in the species is given by $dN/dt \sim \alpha N$, where $\alpha$ represents the birth rate of the individuals (the model is symmetrical with respect to death if the the death rate of individuals is represented as $-\beta$). The solution of the above equation assumes an exponential form, $N \sim \exp(\alpha t)$, which is not a good representative model for population dynamics in real-world bio-systems as discussed above. The indefinite increase in the population is capped by defining a carrying capacity for the system. The differential equation describing this capped population growth model is given by $dN/dt \sim \alpha N(1 - N/N_{max})$, where $N_{max}$ is the maximum carrying capacity for the population the system can support. The solution to above model gives rise to a logistic (S-shaped) curve with $N \sim N_{max}/(1 + N_{max}\exp(-\alpha t))$ and represents a simple description of the population evolution for a single species system. In the case of multiple species that are competing against each other, the competition model uses a system of coupled non-linear differential equations. For two species, $N_1$ and $N_2$, the set of equations are,
\begin{equation}
\label{eq:coupled1}
   dN_{1}/dt\sim\alpha N_1 - \gamma_{12} N_1 N_2
\end{equation}
\begin{equation}
\label{eq:coupled2}
   dN_{2}/dt\sim -\beta N_2 +\gamma_{21} N_1 N_2.
\end{equation}
The inter-species interaction is given by the non-linear cross term and $\gamma_{ij}$ gives the magnitude of interaction. The solutions to the Eqs.~\eqref{eq:coupled1} and~\eqref{eq:coupled2} depend upon the parameters $\alpha$, $\beta$, and $\gamma$. In higher dimensions ($N \geq 5$), such a system of equations have been found to generate asymptotic behaviors that include a fixed point, a limit cycle, an $n$-torus, and/or attractors~\cite{smale1976differential, hirsch1989systems}. However, for large $N$-systems, this Lotka-Volterra-type model predicts instability. Such systems have been observed to exhibit stability if the species evolve by natural selection~\cite{ackland2004stabilization}.

The above approaches to model population dynamics captures only the temporal dynamics of the system. However, the spatial aspects can influence the interactions between species through dispersion and competition for space. 

\section{Reaction-diffusion Model for Population Growth with Convection}
In order to account for the spatial and temporal behaviors, the growth-diffusion model has been proposed where, in addition to the logistic growth model as above, the concept of diffusion is also integrated. In a three or higher dimension realization of this model convection currents also originate as the spatial degrees of freedom are not constrained within the plane. This allows the individuals of a species the spatial freedom to disperse and in turn compete for resources due to nearby occupied neighbors. The governing equation for a growth-diffusion model with convection therefore takes the following form,
\begin{equation}
\label{eq:spatial}
\begin{split}
   dN(\textbf{r}, t)/dt = \alpha N(\textbf{r}, t)(1 - N(\textbf{r}, t)/N_{max})\\
   + \mathcal{D}\nabla^2 N(\textbf{r}, t) + \vec{V}\cdot\nabla N
\end{split}   
\end{equation}
The penultimate term in Eq.~\eqref{eq:spatial} takes into account the diffusion of the individuals in space. The diffusion constant, $\mathcal{D}$, controls the rate of diffusion in space described by ($\textbf{r}\in\mathcal{R}^3$). The final term in Eq.~\eqref{eq:spatial} describes the convective motion of the general convection-diffusion differential equations. The convective element appear as the population hits a sufficiently high concentration to be able to disperse and drift in space. This model takes into account the stochasticity with respect to population growth and deaths, thus, represents a more realistic outlook for the population dynamics in nature. The solution to Eq.~\eqref{eq:spatial} is obtained numerically in a confined volume with each individual of a species designated to make a random walk in the defined region with a well-defined mean and variance.

\section{Experimental Methodology}
To compliment this theoretical development and obtain high-temporal-resolution population evolution of a \textit{E. Coli} culture for comparison, a unique light-scattering experiment was designed. Dynamic light-scattering (DLS) is a technique that can be used to determine many physical characteristics of the scatterers such as the total number, size, and distribution profile of small particles in suspension. In addition, dynamics of the particles in solution may be measured, where the temporal fluctuations are usually analyzed by means of the intensity or photon auto-correlation function (also known as photon correlation spectroscopy or quasi-elastic light scattering). Here, a technique was developed in order to obtain finer time resolution of the population number of a culture of bacteria. The present technique can be extended to obtain all other measurements from a DLS experiment simultaneously and is denoted as \textbf{ARGOS} (area recorded generalization optical scattering) technique. The key features of this technique were the usage of a semi-translucent screen that is illuminated by the scattered light and recorded by a CCD camera imaging the screen and a through-beam filter, made up of multiple layers of neutral-density filters (ND), to calibrate the laser intensity and stability over long experimental runs~\cite{saad, abby}. If the scattering sample is chosen either dilute enough or thin enough such that only single scattering events are likely to occur, then the total integrated intensity of each image, corrected for thru-beam intensity, is proportional to the number of scatters in the scattering volume. For this experiment, since the scattering geometry (placement of the screen to be imaged down range) covers a wide solid angle, the total integrated intensity is a very good measure of the scatter density/population with a precise time stamp.

Population monitoring experiment used a genetically modified \textit{E. Coli} capable of infection by F-specific coliphage, or F-amp, and resistant to antibiotics to ensure a specific population of bacteria. The advantages of using F-amp \textit{E. Coli} bacteria are ready availability and are genetically engineered for specific traits, and, for later experiments, can be easily infected. The F-amp solution was prepared with $1$~mL of log phase bacteria, $0.1$~mL of streptomycin/ampicillin antibiotic solution and $8.9$~mL PBS solution to ensure single scattering events. This sample was placed in a $10 \times 10$~mm cuvette at a temperature of $30 \pm 1$~$^\circ$C, and exposed to the $< 5$~mW laser (He-Ne) for runs lasting about $48$~hours with images collected logarithmically in time from every $1/30$~sec initially to every $5$~min toward the end of the run~\cite{pecora1964doppler, goodman1976some, aragon1976theory}. The thru-beam in the image is integrated for each image and the total image intensity is scaled to that of the first image thru-beam intensity to correct for drift in laser intensity over long runs. The entire scaled image is then integrated outside the center region of the ND filters to the image edge for each image to yield the total image intensity, directly proportional to population of bacteria, as a function of time. Experiments were repeated to confirm reproducibility.

\section{Results and Discussion}
\begin{figure}[t]
\includegraphics[scale=0.5]{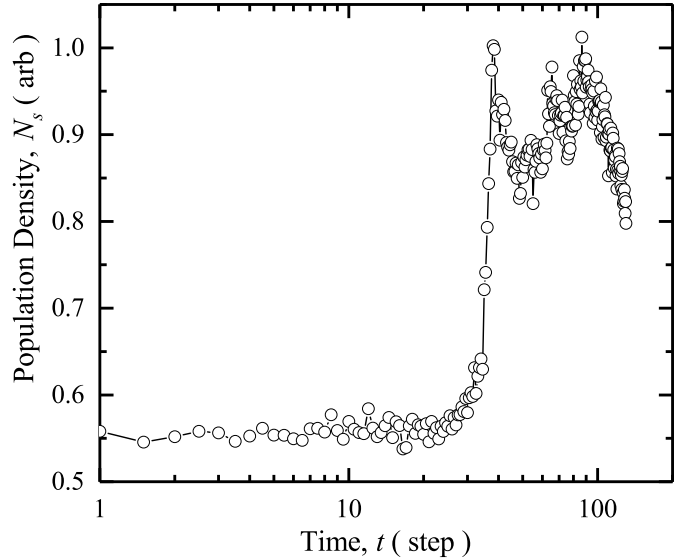}
\caption{ \label{fig1} Figure shows the experimental results with visibly distinguishable phases: lag, growth, and death. The saturation phase is observed to be absent.}
\end{figure}

The life-cycle of a typical bacterial cell culture consists of the four distinct phases: the lag, growth, stationary, and the death phase. Although stochasticity is associated with these phases, the general nature of the curve assumes a plateau-like form in the stationary phase due to the exhaustion of resources. The data obtained from the \textbf{ARGOS} technique at finer scales reveal the expected phases but rather than a flat plateau, the stationary phase exhibited oscillations in population. See Fig.~\ref{fig1}.

\begin{figure}[b]
\includegraphics[scale=0.5]{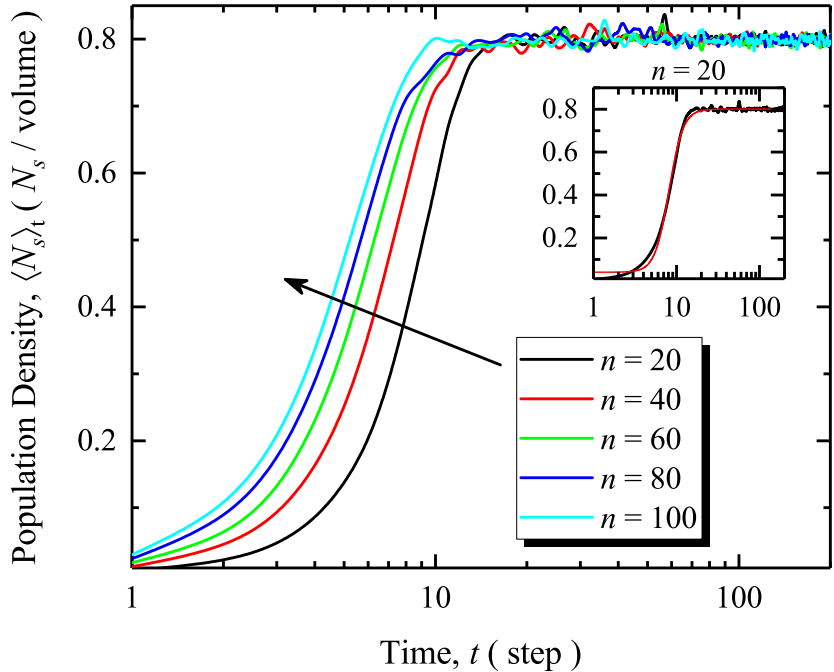}
\caption{ \label{fig2} The numerically solved average population density, $\langle N_s\rangle_t$ as a function of time with varying initial population, $n = 20, 40, 60, 80, 100$ individuals and $\sigma_c = \sigma_d = 0.4$ on a semi-logarithmic scale averaged over $5$ simulation runs each for $200$ time steps. The arrow indicates increasing $n$. Inset plot shows a sigmoidal fit for $n = 20$.}
\end{figure}

The presence of such oscillating populations \textit{in solution} indicate intra-species mixing and competition between groups of individuals in the population that are not spatially homogeneous. This experimental observation is reproducible and has only been hinted at using traditional survey techniques~\cite{Finkel2006} or in bacterial film colonies~\cite{Liu2017}. To simulate the above observations, a three-dimensional ($3-$d) lattice using the spatial logistic model approach given in Eq.~\eqref{eq:spatial} is used~\cite{law2003population}. The intrinsic growth and death rates of individuals are given by the parameters $\alpha$ and $-\beta$. A random seed is chosen that specifies the initial location of an individual in the lattice and, due to their intrinsic growth rate, will multiply and disperse within the lattice region. The dispersal of an individual in the $3-$d lattice is modeled as a random walk. On the discretized $n \times n \times n$ lattice, a random cell ($r_i$) is chosen and its dispersion ($|r_i - r_j|$) is given by a function, $f(r_i, r_j)$. The growth rate ($\mathcal{G}(r_i, r_j, t)$) is then give by $\mathcal{G}(r_i , r_j, t) = \alpha f(r_i, r_j)$. The kernel dispersion, $f(r_i, r_j)$, is taken as a trivariate Gaussian distribution function with a well-defined mean $\mu$ and a variance $\sigma_d$, written as $f(r_i, r_j)\sim\exp(-(r_i -\mu)^2 / 2\sigma_d^2)$. The variance, $\sigma_d$, signifies the magnitude of dispersion on the lattice. Competition for space among the individuals occupying neighboring kernels is then modeled in a similar way. The interaction between the neighboring kernels, $\gamma$, and the intrinsic death rate $-\beta$ is given by $\mathcal{D}(r_i, r_j, t) = -\beta + \Sigma_i\gamma g(r_i, r_j)$. Here, the competing forces and their `range of interaction' are given by another trivariate Gaussian distribution with a well-defined mean $\mu$ and a variance $\sigma_c$ with $g(r_i , r_j)\sim\exp(-(r_i -\mu)^2 / 2\sigma_c ^2)$. The convective element is calculated for the vertical spatial dimension, by taking the derivative of the concentration (along that direction), and then multiplied by the average vertical velocity of each layer.  

\begin{figure}[hb!]
\includegraphics[scale=0.5]{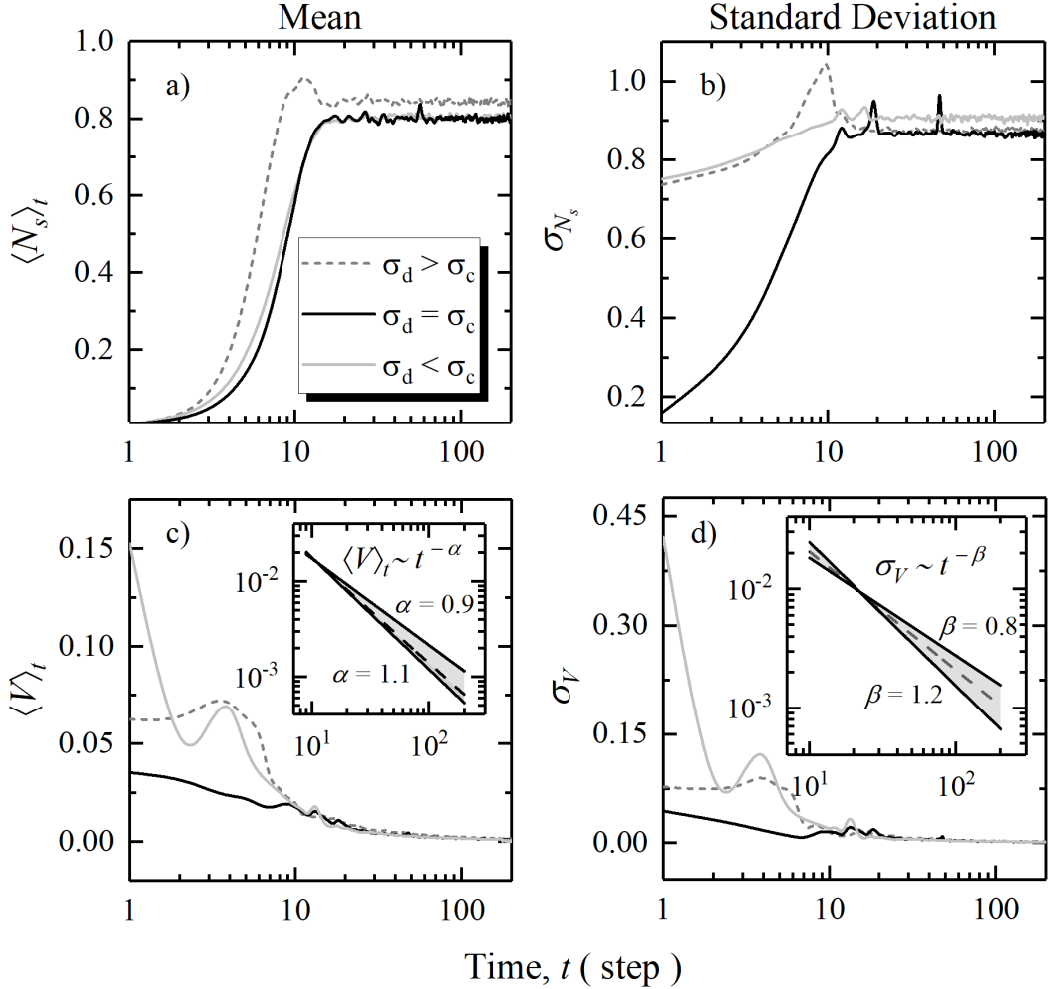}
\caption{ \label{fig3} a) The simulated average population density, $\langle N_s\rangle_t$ with an initial population of $n = 20$ individuals for the three cases: $\sigma_d = 0.6 > \sigma_c$ = 0.4 (dashed black), $\sigma_d = \sigma_c = 0.4$ (light grey), and $\sigma_d = 0.4 < \sigma_c = 0.6$ (solid black) on a semi-logarithmic scale averaged over $5$ simulation runs each for $200$ time steps. b) The standard deviation of the population density, $\sigma_{N_s}$ as a function of time step for the three cases. c) The average convective velocity, $\langle V\rangle_t$ as a function of time. Inset plot shows the functional dependence of the average convective velocity on time on a log-log scale. The average velocities from the three cases converge as, $t^{-\alpha}$, $0.9<\alpha<1.1$, when $t\geq 10$. d) The standard deviation of the velocity, $\sigma_{V}$ as a function of time step for the three cases. Inset plot shows the functional dependence of the standard deviation of the convective velocity on time on a log-log scale, $\sigma_V\sim t^{-\beta}$, $0.8<\beta<1.2$, when $t\geq 10$.}
\end{figure}

Fig.~\ref{fig2} shows the solution of the $3-$d steady-state population number density (in inverse volume) as a function of time using varying initial seeding of $n = 20, 40, 60, 80, 100$ individuals occupying random lattice sites in space with equal dispersal and competition effects, $\sigma_c = \sigma_d$, for $200$ time steps. The simulated population density reveals the lag, growth, and stationary phases but not yet the death phase, as it is not built in the current model. With an increase in initial seeding it is observed that the maximum population density is achieved faster with steeper growth slopes. The inset plot in Figure~\ref{fig2} shows a sigmoidal fit for the initial seeding, $n = 20$.

\begin{figure}[hb]
\includegraphics[scale=0.5]{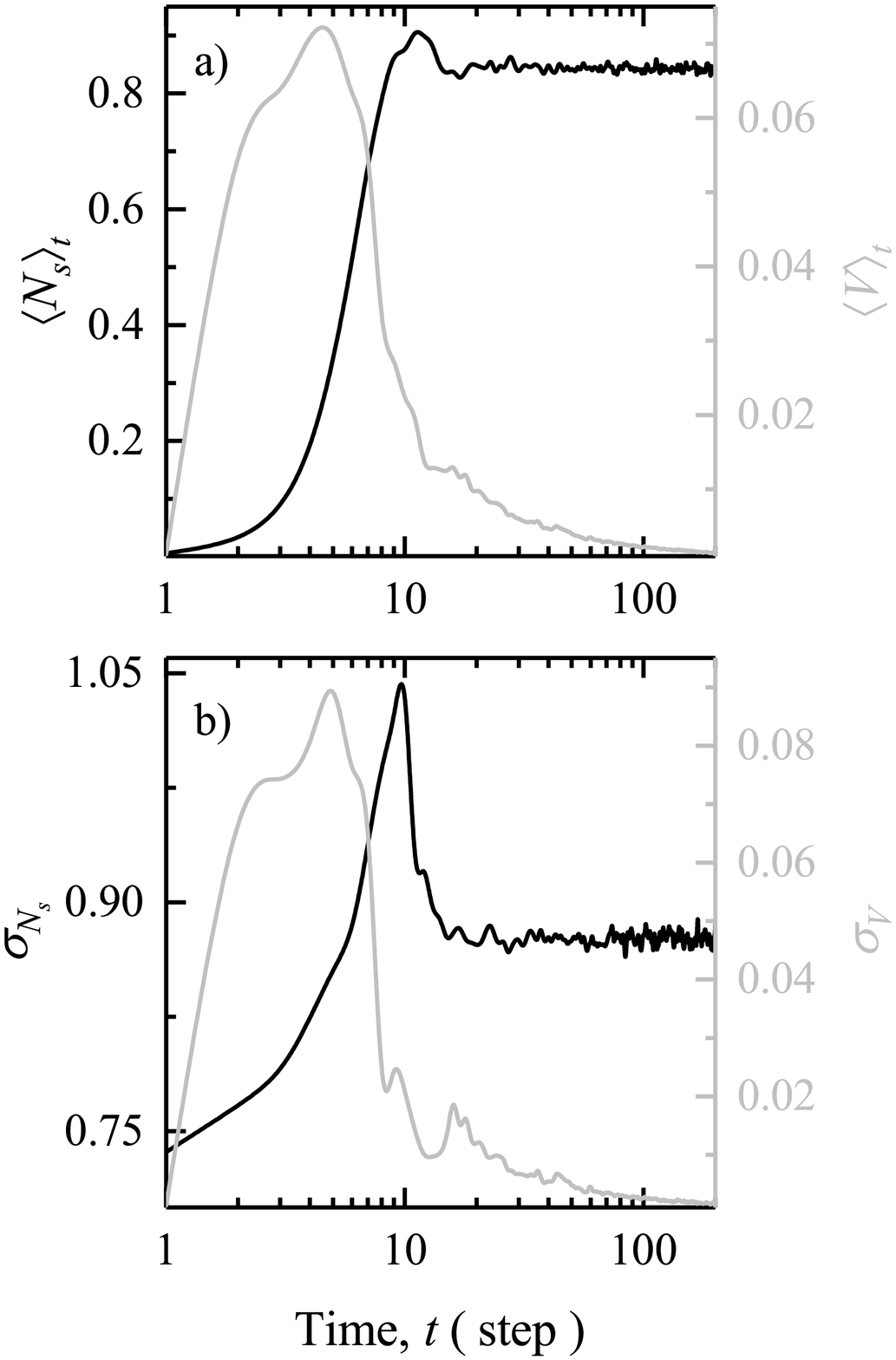}
\caption{a) The simulated mean of the population density (left axis) and the convective velocity (right axis) plotted as a function of time for $\sigma_c = \sigma_d$ on a semi-log scale. Note that the two graphs intersect when population density hits $\sim 70$ $\%$. b) The standard deviation of the population density (left axis) and the convective velocity (right axis) as a function of time under the same model constraints.\label{fig4}}
\end{figure}

To explore the roles of the two effects: competition and dispersion through their respective variances, $\sigma_c$ and $\sigma_d$, multiple simulations were run for three cases; $\sigma_d > \sigma_c$, $\sigma_c = \sigma_d$ and $\sigma_d < \sigma_c$ as shown in Fig.~\ref{fig3}. As seen in Fig.~\ref{fig3}a, an increase in $\sigma_d$ relative to $\sigma_c$ increases the population density over all time steps while substantially shortening the growth phase. However, the steep increase in the population density is compensated by competition for space as can be observed by the presence of a hump as the mean population density hits $\sim 90$ $\%$. The standard deviation of the population density in Fig.~\ref{fig3}b shows a steady increase for the case, $\sigma_c = \sigma_d$ from $0.2$ to upto $0.9$. This signifies the steady dispersion of the agents within the lattice as time proceeds. For the case, $\sigma_d>\sigma_c$, the standard deviation initially increases, and then it collapses once competition starts dominating. From both Fig.~\ref{fig3}a and~\ref{fig3}b it can be seen that at times $t\geq10$ when the mean population density hits a saturation, the dynamics from all the three cases almost converges, and little or no difference is observed thereafter. In panels Fig.~\ref{fig3}c and~\ref{fig3}d we plot the mean and the standard deviation of the convective velocities for the three cases. The mean velocities are observed to decrease as time progresses, with the maximum decrease observed in the case when $\sigma_c>\sigma_d$. However, at $t\geq 10$ the mean velocities from all three cases converge and decay as one. The inset plot shows the allometric relationship between mean convective velocity and time when $t\geq 10$, $\langle V\rangle_t\sim t^{-\alpha}$, where $\alpha = 1.1$ when $\sigma_c>\sigma_d$, $\alpha = 1$ when $\sigma_c = \sigma_d$, and $\alpha = 0.9$ when $\sigma_d>\sigma_c$. In Fig.~\ref{fig3}d we plot the standard deviation of the convective velocity as a function of time. As expected, the case with $\sigma_c>\sigma_d$ shows the largest fluctuation, however the three cases can be collapsed into allometric relationships as can be seen the inset plots, $\sigma_V\sim t^{-\beta}$. We discuss the theoretical details of the observed relationships in detail in Appendix. 

In Fig.~\ref{fig4}a we show the crossover plots for the mean population density and the mean convective velocity for the case when $\sigma_c = \sigma_d$. When the population density hits~$\sim 70\%$ at time close to  $10$ simulation time-steps, the two plots intersect, and thereafter the population density hits saturation while the convective velocity decreases. In Fig.~\ref{fig4}b we can see that the standard deviation in the population density increases over time and peaks at $t=10$, after which it starts decreasing and then becomes constant. A trend like above throws light as to how the population of the agents disperse in the $3$d-lattice in time. The standard deviation of the velocity plot is observed to intersect the standard deviation of the population density at~$t=10$. When $t\leq 10$, the standard deviation of the velocity plot is observed to increase, thus implying random motion both in-plane as well as, out-of-plane. However, when $t>10$ the standard deviation is observed to decrease thus signifying a coordinated motion, like out-of-plane convection currents. By observing both Fig.~\ref{fig4}a and~\ref{fig4}b we can conclude that the convection current is short lived as the convective velocity starts rapidly decreasing as the population density achieves a steady-state. A potential reason for this is over-crowdedness and a lack of enough empty lattice sites.

From these simulations, the driving parameter appears to be the dispersal variance $\sigma_d$ when the motion is confined to a single plane. The convective current is observed to be driven by the competition variance $\sigma_c$. The dispersal variance allows for two-dimensional spatial spreading or in-plane diffusion while competition variance  gives rise to out-of-plane spatial exploration in search of more lattice vacancies. This is understandable as $\sigma_d$ controls the localization of the individuals while $\sigma_c$ controls the width of the competition effect (magnitude) distribution. As the lattice becomes crowded, the stationary phase density is just achieved quicker for a given set of variances. It should be noted that a slight decrease in the population density was observed in the simulations (not shown here) but only for very long run times and much denser initial seeded populations. Since the experimental data were taken only up to the stationary phase, the simulations focused on just the first three phases. While the experimental data are reasonably well represented by the 3-d lattice simulations, the unique population oscillations observed in the stationary phase are not captured.

In this paper, the population dynamics of a relatively dilute solution culture of F-amp \textit{E. Coli} bacteria was measured with high temporal resolution and revealed not only the expected phases but an oscillatory behavior during the stationary phase indicating that the population as a whole is not homogeneous and that competition between sub-groups can play an important role. The simulation model developed here attempts to incorporate dispersion and competition simultaneously between individuals within the populations. The results of the numerical simulations captures the main features of the observed population evolution but not fully the oscillating behavior. Apparently, the present numerical model does not result in sub-group populations acting within the overall population. An expansion of the model in this direction is needed but it is important that the sub-groups emerge naturally from the simulation and not be imposed. Taken together, the experimental and simulation results shed light onto the problem of understanding the evolution of populations and should be of interest to a wide-range of disciplines




\section{Appendix}
We discuss the analytical implications of the stochastic model in this section from an empirical point of view. The rate of change in the average convective velocity is given as the difference in the spatially-averaged velocities of two adjacent layers of the cubic lattice along the $z$-direction at every time-step of the simulation. The system is finite, and as seen from the simulation results in Figures~\ref{fig3} and~\ref{fig4} the average convective velocity decreases with time. Hence we note that, $d\langle V\rangle_t/dt = -\Big(\frac{\langle V\rangle_t|_{z+1} - \langle V\rangle_t|_{z}}{t_s}\Big) = -\frac{1}{t_s}\Big(\frac{\delta h_+}{\delta t}|_{z+1} - \frac{\delta h_+}{\delta t}|_{z}\Big)$. Here, $\delta h_+$ denotes the positive upward draft of the agents from a lattice layer, $\delta h_+ = |z_{max} - z|$ where $z\leq z_{max}\leq z+1$ is the maximum upward draft between $z$ and $z+1$. The driving force for the convective flow lies in the interplay between competition and diffusion primarily competition, as diffusion allows for two-dimensional spatial spreading while competition gives rise to out-of-plane spatial exploration. Therefore, we make an assumption that the net upward draft on an average is, $\delta h_+ = (\sigma_c/\sigma_d)^n\langle z\rangle$ for some arbitrary real `$n$' with respect to $\langle z\rangle$. The time-rate change of the average convective velocity then becomes, $d\langle V\rangle_t/dt = -\frac{1}{t_s}(\sigma_c/\sigma_d)^n\Big(\langle\frac{z+1}{t}\rangle - \langle\frac{z}{t}\rangle\Big) = -(\sigma_c/\sigma_d)^n(\langle V\rangle_t/t)$. Solving the differential equation we get, $\langle V\rangle_t\sim t^{-(\sigma_c/\sigma_d)^n}$. 
The ratio, $\sigma_c/\sigma_d$ is \emph{not} allowed to exceed $2$, as competition overpowers growth and dispersion thus terminating the simulation soon after the initial random seeding. The ratio of the two variances when expressed in a general form, $(\sigma_c/\sigma_d)^n < 2$, must simultaneously satisfy the constraint, $0<(\sigma_c , \sigma_d)<1$. The lower bound however is not explored in the current scope of study.

We define an arbitrary velocity distribution function, $f(t)$ such that, $\langle V\rangle_t = \int tf(t)dt$. Let $(\sigma_c/\sigma_d)^n = \alpha$ then, $t^{-\alpha} = \int tf(t)dt$, and the functional differential equation becomes, $tf^{\prime}(t) + f(t) + \alpha t^{-(\alpha + 1)} = 0$. On solving, $f(t) = c/t + t^{-(\alpha + 1)}$, where $c$ is an arbitrary constant. As the distribution function has to be normalizable, $\int_0^\infty dt f(t) = 1$, the constant, $c=0$. The standard deviation is then given by, $\sigma_V (t) = \int (t - t^{-\alpha})^2 f(t)dt = -t^{-(\alpha-2)}/(\alpha-2) - t^{-3\alpha}/3\alpha + 2t^{-(2\alpha-1)}/(2\alpha - 1)$. 
For $\alpha > 2$ the standard deviation is defined over the infinite domain however, for $\alpha >1/2$ the standard deviation is undefined. We saw earlier that $\alpha$ is bounded ($\alpha < 2$) therefore, we consider the case when $(2\alpha - 1)>0$. It is clear that the variance diverges off to infinity when defined over the infinite domain. Thus, we identify the second exponent, $\beta = 2\alpha - 1$. As $1/2<\alpha<2$, $n\leq 1$ satisfies all the imposed constraints for all ratios of variances under constraints. We define a positive integer $d$ such that, $n=1/d$ and observe the slow convergence in the values of the exponents $\alpha$ and $\beta$ when compared with the simulation fits. Thus, the analytical expressions for the simulation results can be explained through the general form, $\langle V\rangle_t\sim t^{-(\sigma_c/\sigma_d)^{1/d}}$ while $\sigma_V (t)\sim t^{-(\sigma_c/\sigma_d)^{(2/d) - 1}}$.



\end{document}